\documentclass[12pt,preprint]{aastex}

\shorttitle{VLBA Measurement of Nine Pulsar Parallaxes}

\received{2001 December 3}
\begin{document}

\title{VLBA Measurement of Nine Pulsar Parallaxes}

\author{Walter F. Brisken}
	\affil{National Radio Astronomy Observatory}
	\affil{PO Box O, Socorro, NM 87801}
	\email{wbrisken@nrao.edu}
\author{John M. Benson}
	\affil{National Radio Astronomy Observatory}
	\affil{PO Box O, Socorro, NM 87801}
	\email{jbenson@nrao.edu}
\author{W. M. Goss}
	\affil{National Radio Astronomy Observatory}
	\affil{PO Box O, Socorro, NM 87801}
	\email{mgoss@nrao.edu}
\and
\author{S. E. Thorsett}
	\affil{Department of Astronomy and Astrophysics, University of California}
	\affil{Santa Cruz, CA 95064}
	\email{thorsett@ucolick.org}

\begin{abstract}
  
  We determined the distances to nine pulsars by parallax 
  measurements using the NRAO Very Long Baseline Array, doubling
  the number of pulsars with accurate distance measurements. Broadband
  phase modeling was used to calibrate the varying dispersive effects
  of the ionosphere and remove the resulting phase errors from the
  phase-referenced VLBI data.  The resulting parallaxes have a typical
  accuracy of 100~microarcseconds or better, yielding distances
  measurements as accurate as 2\%.  We also report new proper motion
  measurements of these pulsars, accurate to
  400~$\mu$as~yr$^{-1}$ or better.

\end{abstract}

\keywords{astrometry --- techniques: interferometric --- pulsars}

\section{Introduction}

Accurate pulsar distance measurements are useful for many reasons.
Combined with dispersion measures, they probe the electron content
along different lines of sight through the local Galaxy.  Combined
with proper motions, they constrain the pulsar velocity distribution
and hence the symmetry of supernova explosions.  Combined with
radio, optical, x ray, and UV flux measurements, they determine the 
absolute luminosities and radiation efficiencies of pulsars.  For a
thermally emitting neutron star, they even promise the possibility of
directly constraining the neutron star radius and hence the nuclear
equation of state at high densities.  There are many different ways of
estimating pulsar distances, including association with supernova
remnants or globular clusters at known distance, or the use of a
model of the electron distribution in the Galaxy together with
dispersion measurements.  But the most fundamental method---indeed,
the only model-independent method---remains the measurement of parallaxes.

Pulsar parallax measurements are difficult.  Even the closest pulsars are
100~pc or more distant; thus parallax leads to an apparent annual
motion of 10~mas or (typically) much less.  With their steep spectra,
pulsars are usually observable only in the low-frequency radio band,
where the ionosphere introduces substantial temporally- and
geographically-variable shifts in the phase of the signal, making
precision interferometry particularly challenging.

Before the completion of the Very Long Baseline Array (VLBA), only six
interferometric pulsar parallaxes were published
(Table~\ref{tab:early}).  It is important to understand the very great
difficulties these observers faced.  First, in many cases they were
limited to rather short baselines ($<500$~km) and decimeter and longer
wavelengths ($\lambda>10$~cm).  The ionosphere and the small array
size limited the potential accuracy to about 1~milliarcsecond (mas),
unless a very nearby calibrator source was used.  Some serendipitous
measurements were possible.  For example, the measurement of the
parallax of pulsar B1451$-$68 was about ten times more precise
than the ionosphere would ordinarily have allowed with the 
available array because of the
fortuitous presence of a calibrator only $6^{\prime}$ from the pulsar,
within the same primary telescope beam.

A second problem for early observers was the limited number of
telescopes available in early, ad hoc VLBI networks.  With only
three or fewer stations, it is very difficult to fully resolve
``phase-wrap'' ambiguities to determine which of a family of possible
positions on the sky is actually occupied by the source.  As shown in
Brisken {\it et al.} (2000), an incorrect choice of phase wraps could
lead to parallax errors much larger than the formal measurement
errors. (By the time the last measurement in Table~\ref{tab:early} was
made---the parallax of B2021+51---a number of VLBA stations was
available. Although the observations used a different set of
telescopes at each observing epoch, and large baseline errors in the
telescope positions had still not been resolved, it is notable that
this early result is within $2\sigma$ of our new measurement.)  It is
probably not coincidental that every repeated pulsar parallax
measurement has yielded a smaller parallax (and greater pulsar
distance).  We suspect that the expectation of larger results in early
experiments probably contributed to incorrect lobe choices.  With the
ten-station VLBA this is no longer a problem.

Although we will not further discuss non-radio measurements here, for
completeness we note that the high-energy pulsar Geminga (J0633+1746)
has a three-epoch parallax measured by HST of $\pi = 6\pm2$~mas
(Caraveo et al., 1996), and the radio quiet neutron star
J185635$-$3754 has an HST-derived parallax of $16.5\pm2.3$ mas
(Walter, 2001).

\clearpage
\begin{deluxetable}{lcccccc}
\tablewidth{0pt}
\tablecaption{Early interferometric pulsar parallaxes \label{tab:early}}
\tablehead{ 
    \colhead{Pulsar}
  & \colhead{Stations}
  & \colhead{Baseline}
  & \colhead{$\nu_{\mathrm{obs}}$}
  & \colhead{epochs}
  & \colhead{$\pi$}
  & \colhead{Reference}
 \\
  &
  & \colhead{(km)}
  & \colhead{(MHz)}
  &
  & \colhead{(mas)}
  &
}
\startdata
B0823+26 & 3    & 5300        & 1660 & 4        & $2.8\pm0.6$   & 1 \\
B0950+08 & 3    & 5300        & 1660 & 4        & $7.9\pm0.8$   & 1 \\
B1451-68 & 2    & 275         & 1600 & 9        & $2.2\pm0.3$   & 2 \\
B1929+10 & 2    & 127         & 480  & $\sim 7$ & $21.5\pm8.0$  & 3 \\
B1929+10 & 4    & 35          & 2695 & 5        & $<4$          & 4 \\
B2021+51 & 4--6 & $\sim 6000$ & 2218 & 4        & $0.95\pm0.37$ & 5 \\
\enddata
\tablerefs{
(1) Gwinn et al., 1986;
(2) Bailes et al., 1990;
(3) Salter et al., 1979;
(4) Backer \& Sramek, 1982;
(5) Campbell et al., 1996
}
\end{deluxetable}
\clearpage

\section{Objectives and Target Selection}

Our successful measurement (Brisken et al. 2000) of the parallax of
PSR~B0950+08, using a new calibration technique to remove the
disruptive effects of the ionosphere from the interferometry data,
prompted us to expand our program of parallax measurements to a larger
sample.  We had two primary goals.  One was the immediate desire to
increase the number of pulsars with well-measured distances.  But a
second goal was to improve our understanding of the new calibration
technique, to test its limitations, and to measure the level of
residual systematic errors.  For this reason, our target selection was
not yet optimized solely to produce an unbiased pulsar sample for
statistical work on pulsar distances.  Instead, we chose nine pulsars
at a variety of declinations, of varying flux densities, and with varying
pulsar-calibrator angular separations.  We also continued our
observations of PSR~B0950+08, to produce an extremely over-constrained
data set for studies of our errors.

An important constraint on target selection was the need for a high
flux density at 1400~MHz.  Based on our experience with B0950+08, we
expected that we would be able to successfully do an ionospheric
correction for pulsars with flux densities above about 10~mJy at
1400~MHz.  We also preferentially targeted pulsars with estimated
distances (from the dispersion model) of less than about 2~kpc.  For
this experiment, we only considered candidate pulsars whose nearest
VLBA calibrator source was no more than $5^{\circ}$ from the pulsar.
Larger separations lead to larger differential ionospheric and
tropospheric effects, and will be more difficult to calibrate.  The
$5^{\circ}$ limit was set after experiments in which we attempted to
calibrate PSR~B0950+08 against the International Celestial Reference 
Frame calibrator source 1004+141, at a separation of $6.9^{\circ}$. 
At this separation the differential propagation effects were too
large to correct using the new calibration technique.
We also rejected pulsars at declinations
below $-10^{\circ}$, where high elevation ($>20^{\circ}$) observations
cannot be made simultaneously at all VLBA stations.  The pulsars
observed are listed in Table~\ref{tab:pulsars}, together with their
dispersion measures and flux densities from the Princeton pulsar
catalog (Taylor et al., 1993).

\clearpage
\begin{deluxetable}{lccccr}
\tablewidth{0pt}
\tablecaption{Pulsars observed \label{tab:pulsars}}
\tablehead{
    \colhead{Pulsar}
  & \colhead{DM}
  & \colhead{$S_{1400}$}
  & \colhead{Gate gain$^{\dagger}$}
  & \colhead{Calibrator}
  & \colhead{$\theta_{\mathrm{sep}}$}
 \\
  & \colhead{(pc cm$^{-3}$)}
  & \colhead{(mJy)}
  & \colhead{(mJy)}
  & 
  & 
}
\startdata
  B0329+54   & 26.776 & 203 & 4.08 & J0302+5331 & $4^{\circ}37^{\prime}$ \\
  B0809+74   & 5.7513 & 10  & 3.27 & J0808+7315 & $1^{\circ}19^{\prime}$ \\
  B0950+08   & 2.9702 & 84  & 3.09 & J0946+1017 & $2^{\circ}52^{\prime}$ \\
  B1133+16   & 4.8471 & 32  & 5.98 & J1143+1834 & $3^{\circ}15^{\prime}$ \\
  B1237+25$^{\ddag}$
             & 9.2755 & 20  & 5.13 & J1240+2405 & $0^{\circ}51^{\prime}$ \\
             &        &     &      & J1230+2518 & $2^{\circ}11^{\prime}$ \\
  B1929+10   & 3.176  & 41  & 4.03 & J1945+0952 & $2^{\circ}20^{\prime}$ \\
  B2016+28   & 14.176 & 30  & 4.99 & J2020+2826 & $0^{\circ}38^{\prime}$ \\
  B2020+28   & 24.62  & 38  & 4.69 & J2020+2826 & $0^{\circ}36^{\prime}$ \\
  B2021+51   & 22.580 & 27  & 4.90 & J2025+5028 & $1^{\circ}30^{\prime}$ \\
  J2145$-$0750& 9.000  & 10  & 1.91 & J2142$-$0437& $3^{\circ}18^{\prime}$ \\
\enddata
\tablenotetext{\dagger}{The improvement in signal-to-noise by not
correlating during off-pulse.}
\tablenotetext{\ddag}{B1237+25 was observed with two reference sources.}
\end{deluxetable}

\clearpage
        
\begin{deluxetable}{lcccc}
\tablewidth{0pt}
\tablecaption{Calibrator positions and flux densities$^a$ \label{tab:calibs}}

\tablehead{
    \colhead{Name}
  & \colhead{$\alpha$(J2000)$^b$}
  & \colhead{$\delta$(J2000)}
  & \colhead{$S_{2267}^c$}
  & \colhead{$S_{8337}^d$}
\\
  & 
  &
  & \colhead{(mJy)}
  & \colhead{(mJy)}
}
\startdata
J0302+5331   & $03^h02^m22^s.7354$ &  $53^{\circ}31'46''.534$ & 200 & 192 \\
J0808+7315   & $08^h08^m16^s.4918$ &  $73^{\circ}15'11''.980$ & 220 & 269 \\
J0946+1017   & $09^h46^m35^s.0693$ &  $10^{\circ}17'06''.126$ & 390 & 248 \\
J1143+1834   & $11^h43^m26^s.0706$ &  $18^{\circ}34'38''.375$ & 230 & 113 \\
J1230+2518   & $12^h30^m14^s.0912$ &  $25^{\circ}18'07''.139$ & 150 & 175 \\
J1240+2405   & $12^h40^m47^s.9870$ &  $24^{\circ}05'14''.184$ & 160 & 92 \\
J1945+0952   & $19^h45^m15^s.9229$ &  $09^{\circ}52'59''.576$ & 250 & 198 \\
J2020+2826   & $20^h20^m45^s.8719$ &  $28^{\circ}26'59''.205$ & 90  & 18 \\
J2025+5028   & $20^h25^m24^s.9726$ &  $50^{\circ}28'39''.550$ & 280 & 165 \\
J2142$-$0437 & $21^h42^m36^s.8999$ & $-04^{\circ}37'43''.518$ & 65  & 33 \\
\enddata
\tablenotetext{a}{All values listed are from the VLBA Calibrator List 
(Beasley et al., 2002).}
\tablenotetext{b}{Source positions are accurate to 15~mas.  Better positions
for some of these calibrators are now known.}
\tablenotetext{c}{Correlated flux density on 8000~km
baseline at 2267~MHz.}
\tablenotetext{d}{Peak source brightness of VLBA image made at 8337~MHz.}
\end{deluxetable}
\clearpage

\section{Observations}

Five epochs of observations for each pulsar were planned over the
course of one year, enough to redundantly determine the proper motion
and parallax even with a failed epoch.  To maximize our sensitivity to
the angular signature of parallax, the measurements were scheduled at
times when the parallactic displacement was at a maximum or minimum.
Observations hence occurred at roughly three month intervals, although
the exact dates varied with telescope scheduling.  Pulsar
J2145$-$0750 was not detected in either of its first two epochs.
This non-detection 
is most likely because of the extreme, long-term scintillation
modulation that affects the signal from this pulsar.  It was dropped
from the program, and the time allotted to other pulsars was increased
slightly.  About ninety minutes of on-source integration was obtained
for each pulsar at each epoch.

In order to make relative position measurements, each pulsar
observation was interleaved with observations of a VLBA calibrator
source with position known to better than 15~mas.  The antennas nodded
back and forth between the pulsar and calibrator with a cycle time of
about five minutes.  This interval is long enough to detect calibrator
fringes yet short enough to allow unambiguous phase connection between
calibrator observations.  In the case of B1237+25 two calibrators were
observed in a cycle, integrating for 100 seconds on the
pulsar and 60 seconds on each calibrator.

All observations were made with the ten-station NRAO VLBA.  The 20~cm
band (1400 to 1740~MHz) was chosen as a compromise between falling
flux densities and increasing resolution as frequency increases.  This
band also offers the wide fractional bandwidths needed for the
ionosphere calibration.  The 338~MHz wide band available from the receivers
was Nyquist sampled in eight 8~MHz spectral windows beginning at 1404,
1414, 1434, 1474, 1634, 1694, 1724 and 1734~MHz.  Pulsar gating was
employed to increase the signal-to-noise of the pulsar measurements by
disabling correlation during off-pulse.  The gate gain listed in Table
\ref{tab:pulsars} is the expected signal-to-noise increase factor
based on the pulse profiles measured at 1400~MHz.  The timing data
needed to construct the pulse arrival time ephemerides was collected
by Andrew Lyne at Jodrell Bank.

\section{Data Reduction}

\subsection{Calibration}

Data reduction was performed with the AIPS software package.
Phase-referencing (Shapiro et al. 1979) was used to relate the
pulsar's position to that of its calibrator.  Corrections for the
structure of the calibrator source were made.  
The absolute position shift caused by structure is less than 1~mas for
most VLBA calibrators.
Epoch-to-epoch position shift caused by differences in the
baseline coverage is a smaller effect but is more important for
relative astrometry.  Phase-referencing
cancels out most of the pulsar's phase errors such as those due to
clock and baseline offsets and bulk propagation effects; however,
propagation effects that are path-dependent are not exactly cancelled
and dominate the remaining phase errors.  At 20~cm, the largest of
these effects is due to the ionosphere.  A technique for removing the
effect of the ionosphere without independent calibration information
was described by Brisken et al.\ (2000).  It is briefly described
here.

The phase shift at frequency $\nu$ due to a free electron column
density of $\mathrm{\sigma_{e^-}}$ is
\begin{equation}
\mathrm{\Delta \phi_{\nu}^{Iono} \propto \frac{\sigma_{e^-}}{\nu}}.
\end{equation}
The wet troposphere also causes phase shifts, but with a non-dispersive
dependence on $\nu$,
\begin{equation}
\mathrm{\Delta \phi_{\nu}^{Tropo} \propto \sigma_{H_2O} \, \nu},
\end{equation}
where $\mathrm{\sigma_{H_2O}}$ is the column density of water in the
atmosphere.  Only the differential phase shifts at two antennas enter
into the visibility phases.  After phase-referencing, only the
difference in visibility phase shifts due to propagation between the
pulsar and calibrator remain.  While the absolute phase shifts may be
several thousand degrees, only the double-differences of their values
are relevant to the position measurements.  For the troposphere, we
find that this amount is typically less than $45^{\circ}$ of phase for
calibrator-target separations of up to $5^{\circ}$ at 20~cm.  In
contrast, the differential ionosphere can be up to $1200^{\circ}$ on
long baselines.

In addition to the unwanted phase shifts due to propagation effects,
the geometric phase associated with the position of the pulsar (the
quantity that is to be determined) also contributes to the pulsar's
visibility phase.  The point-like nature of pulsars simplifies the
geometric phase to
\begin{equation}
\phi_{\nu}^{\mathrm{Geom}} = \frac{\nu}{c}(\ell u + m v),
\end{equation}
where $(\ell, m)$ are the pulsar's offset from the correlation center
in radians, $c$ is the speed of light, and $(u, v)$ are the components
of the projected baseline perpendicular to the direction to the
pulsar.  The phase delays due to geometry and the troposphere are both
independent of frequency (non-dispersive), making their
disentanglement impossible for a given visibility.  Visibility 
measurements spanning 
a large range of elevations can be used to separate out the troposphere,
but this was not done due to the limited elevation ranges that we used.
Fortunately, it is possible to measure
the differential ionosphere strength with multi-frequency data since
the ionosphere's phase delay is frequency dependent (dispersive).  The
visibility phase on a single baseline can be expressed as
\begin{equation} \label{eqn:fit}
\phi_{\nu}^{\mathrm{Vis}} = A \nu + \frac{B}{\nu} + 2 \pi n.
\end{equation}
where $A$ incorporates all of the non-dispersive components (the
geometry and the troposphere), $B$ is the strength of the differential
ionosphere, and the integer $n$ is the number of additional phase
wraps.  Once $B$ is known, the distorting effects due to the
ionosphere can be removed from the visibility.

Station-based ionosphere solutions were determined by fitting
station-based phases to Eqn~\ref{eqn:fit}.  The station-based phases
at each frequency band were determined by fitting the measured phases to 
a point source model representing the pulsar with solution intervals
shorter than the amount of time that the ionospheric phase shift
changes by about $30^{\circ}$ (about 20 or 30~seconds at 20~cm).
Solution intervals of 10~seconds or less were used on the brightest
pulsars.  Since all phases are relative, one station was chosen to be
the reference antenna and was assigned zero phase.  

\begin{figure}
\plotone{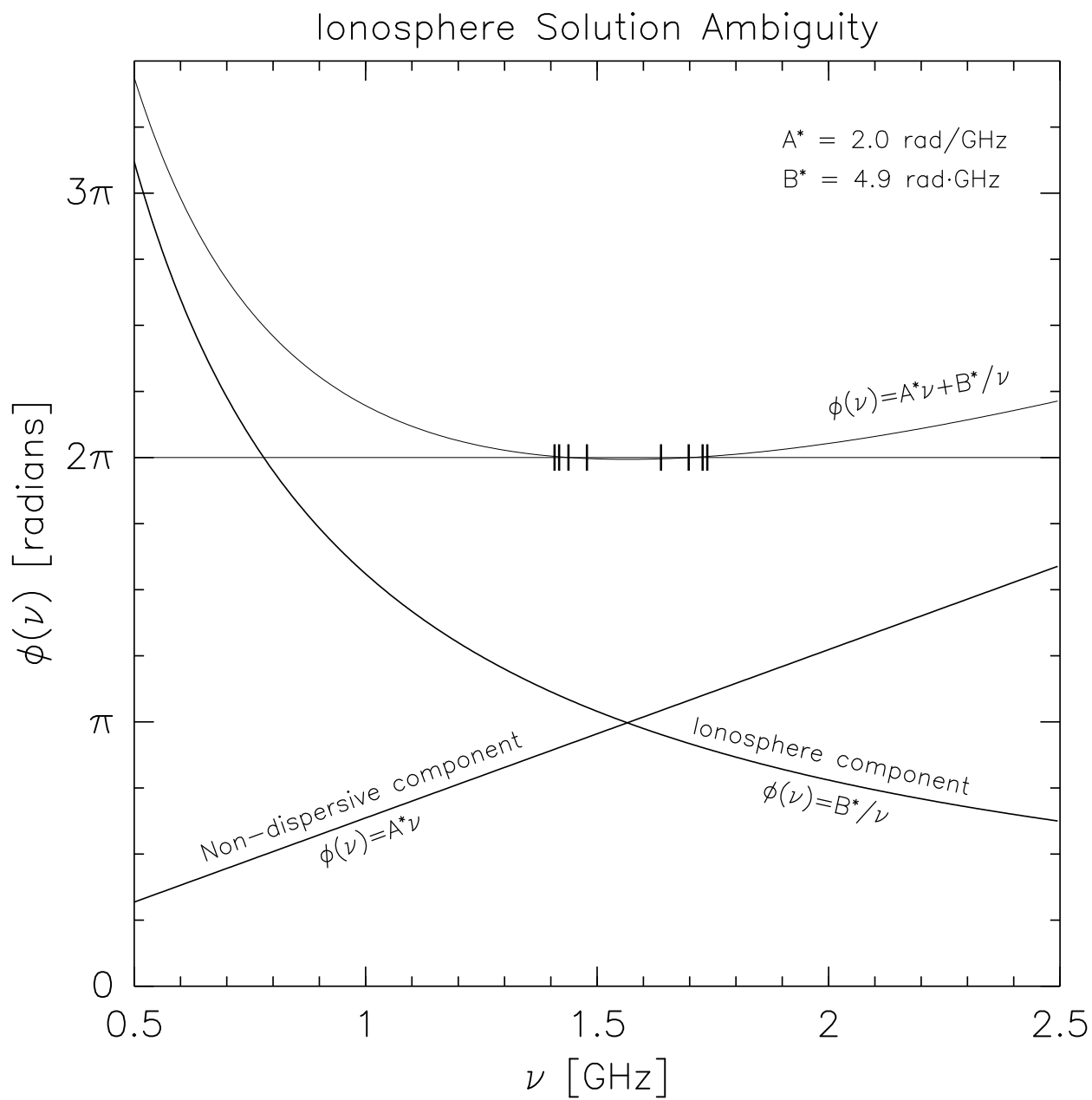}
\caption[ambig.eps]{
\label{fig:ambig}
The source of phase-wrap ambiguity.  For certain ionosphere solutions,
$A^* = 2.0$~rad/GHz and $B^* = 4.9$~rad$\cdot$GHz, the resultant phase
across the observed band is almost indistinguishable from $2 \pi$.
Since $2 \pi$ can be added to the visibility phases without observable
consequences, an entire family of ionosphere solutions is consistent
with the data.  The eight 8~MHz bands observed are shown as short
vertical bars.  The RMS deviation from $2 \pi$ at these frequencies is
only $0.6^{\circ}$.  }
\end{figure}

Unknown phase wraps complicate the fitting for $A$ and $B$.  This is
because certain combinations of the $A$ and $B$ parameters closely
mimic an additional undetectable phase wrap due to the limited
spanned bandwidth as is shown in Figure~\ref{fig:ambig}.  The fitting
was performed first on the Southwestern Antennas (Fort Davis, TX, Kitt
Peak, AZ, Los Alamos, NM, and Pie Town, NM) where baselines are
shorter and thus the geometric phases are smaller and the differential
propagation effects are also less.  An image was made without any
ionosphere calibration to determine the pulsar's position to better
than 10~mas.  The pulsar's approximate position was used to reduce the
unknown non-dispersive phase to less than $90^{\circ}$.  Since the
tropospheric phase is usually less than $45^{\circ}$, the number of
phase shifts, $n$, can be determined and the ionosphere can be
calibrated away for the Southwestern antennas.  The phase shifts
associated with the fitted $B$ parameters were removed from the visibility
data.  An image made with the ionosphere calibrated Southwestern
antennas yielded pulsar positions to better than 1~mas, allowing
ionosphere solutions to be found at and applied to the remaining
stations.

\clearpage

\clearpage

\begin{figure}
\plotone{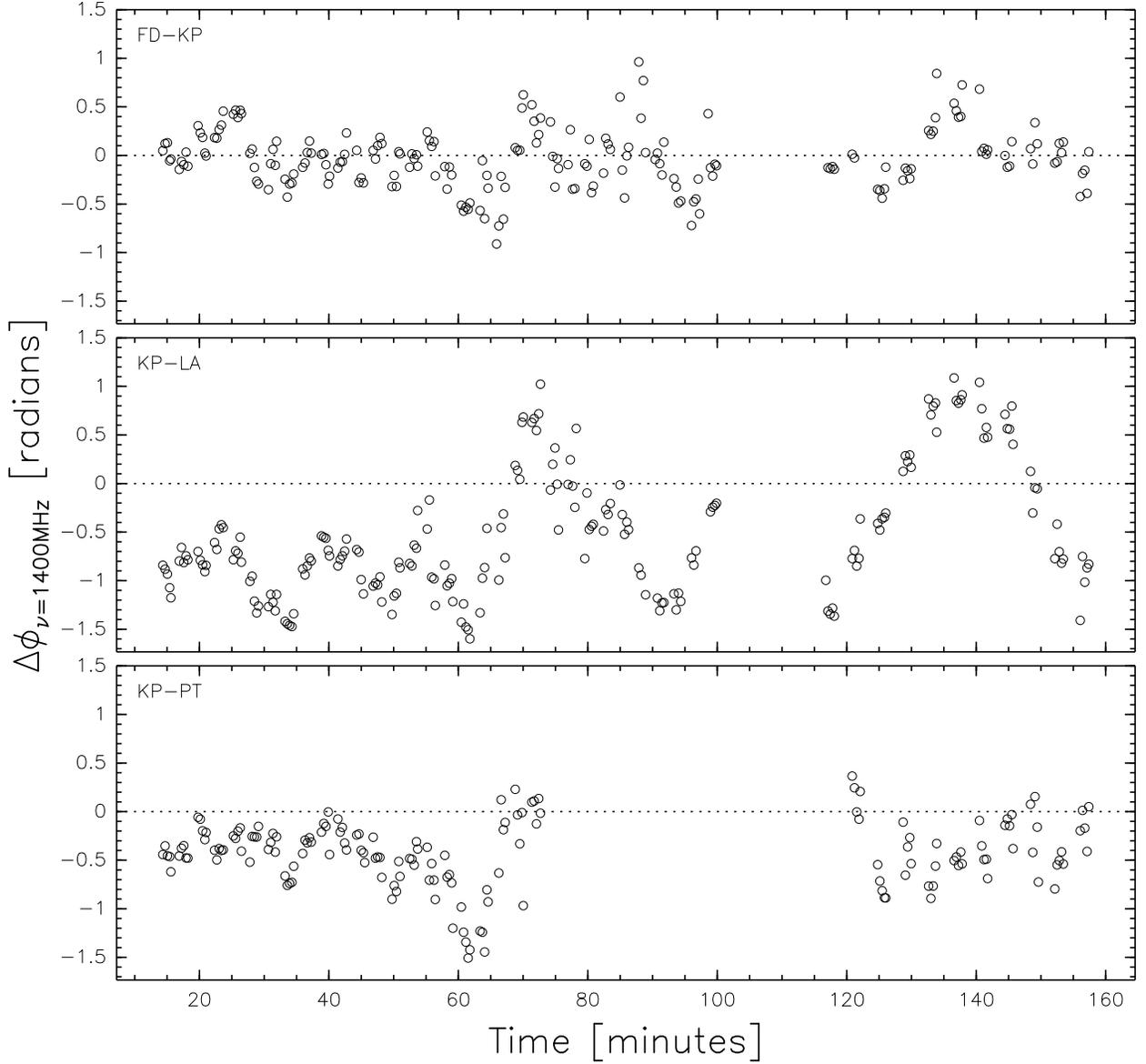}
\caption[ionoshort.eps]{
\label{fig:ionoshort}
The relative phase due to the different residual ionospheric content
above two antennas that is left uncalibrated by phase-referencing.
Kitt Peak (KP) was used as the reference antenna in this observation.
The three shortest baselines to Kitt Peak are shown above.  Distances
are: Fort Davis (FD) to KP = 744~km, KP to Los Alamos (LA) = 652~km,
and KP to Pie Town (PT) = 417~km.  These solutions are from the
phase-referencing of PSR~B1929+10 via calibrator J1945+0952, a
separation of $2^{\circ}20^{\prime}$.  The gap in the data between
100 and 115~minutes is due to the observation of the bandpass calibrator.
The larger gap on the KP--PT baseline is due to telescope failure at
PT.  The date of this observation was 1999 November 16. }
\end{figure}

\clearpage

\begin{figure}
\plotone{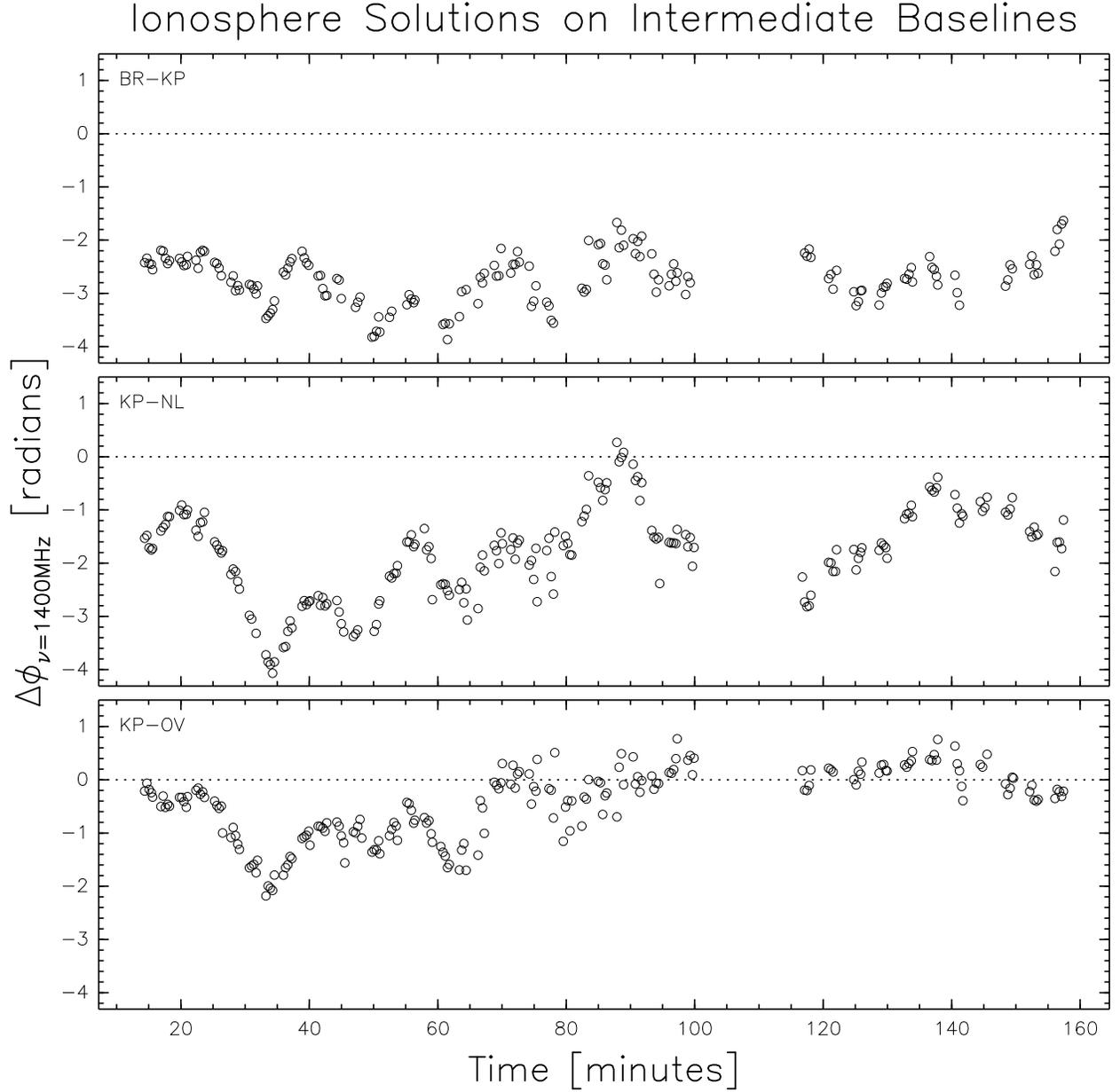}
\caption[ionomed.eps]{
\label{fig:ionomed}
Same as Figure~\ref{fig:ionoshort} but on the three intermediate length
baselines to Kitt Peak.  
Distances are: Brewster (BR) to KP = 1913~km, KP to North Liberty
(NL) = 2075~km, and KP to Owens Valley (OV) = 845~km.  The bias in each
frame is due
to systematic differences in the line-of-sight electron density at the
two stations.
}
\end{figure}

\clearpage

\begin{figure}
\plotone{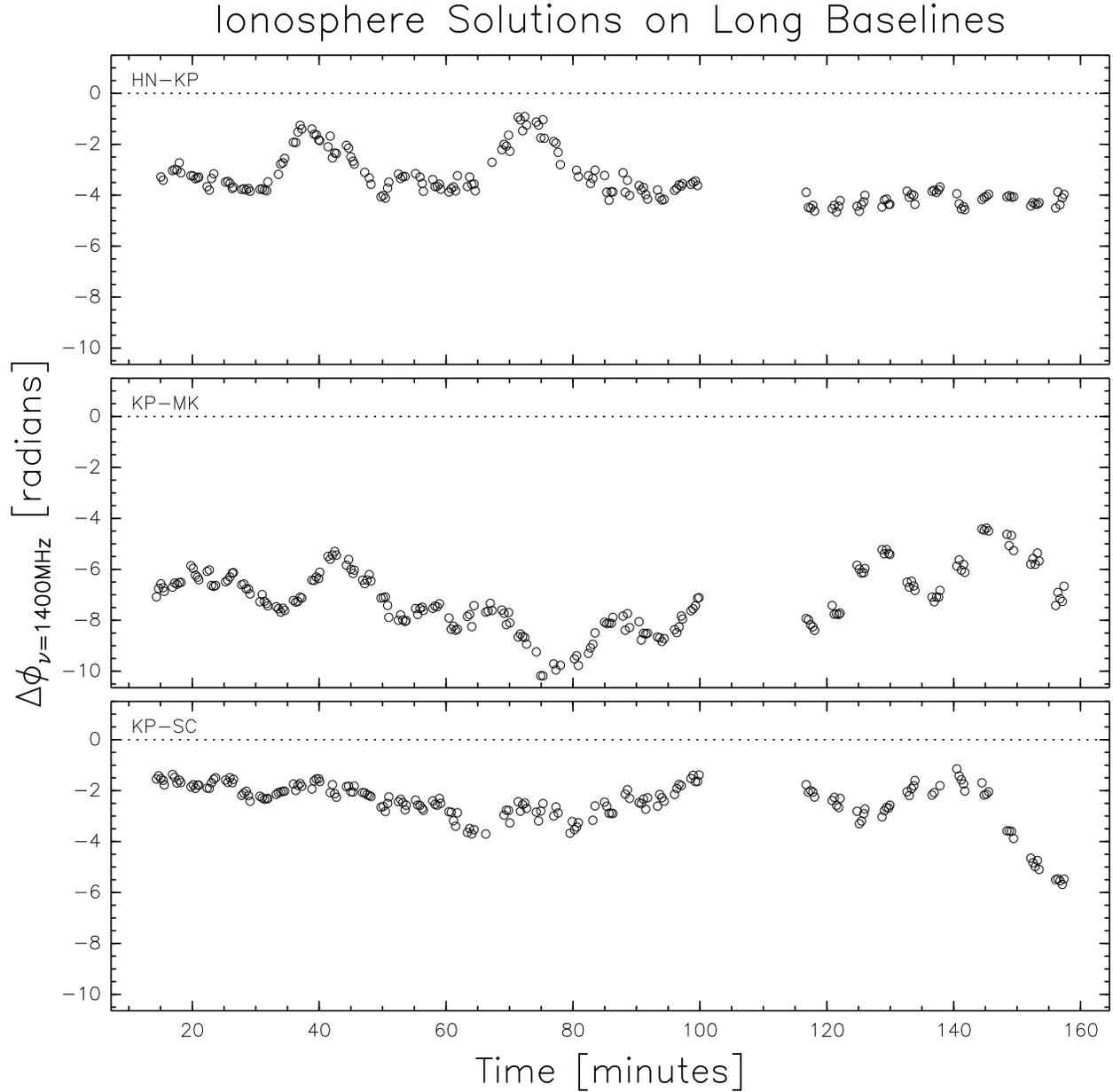}
\caption[ionolong.eps]{
\label{fig:ionolong}
Same as Figures~\ref{fig:ionoshort} and \ref{fig:ionomed}, but on the three
longest baselines to Kitt Peak.  
Distances are: Hancock (HN) to KP = 3623~km, KP to Mauna Kea
(MK) = 4466~km, and KP to Saint Croix (SC) = 4839~km.  For these longer 
baselines the systematic bias is greater.
}
\end{figure}

\clearpage

Differential ionosphere strengths, converted to their effect on
visibility phases at 1400~MHz, are shown as a function of time on all
baselines to Kitt Peak, the reference antenna for this observation, in
Figures \ref{fig:ionoshort}, \ref{fig:ionomed}, and
\ref{fig:ionolong}.  The phase differences on the four shortest
baselines shown (KP to FD, LA, OV, or PT) all have RMS values less
than 1 radian, and are centered close to zero phase.  The longer
baselines often show large systematic offsets from zero phase.  See
for example the BR--KP and KP--MK baselines.  Without correction a
displacement of the pulsar from its true sky location results.  The
data in these figures is from the 1999 November 16, 
observations of PSR~B1929+10.  
The sun was 68$^{\circ}$ from the pulsar during this
observation.  Sunset occurred at Saint Croix during the observation and
is probably responsible for the rapid change in ionospheric phase
delay near the end of the observation on the KP--SC baseline.
Observations made when the pulsar and sun are on opposite sides of
the Earth show considerably smaller ionospheric phase delays but still
exhibit fluctuations on the 10~minute timescale.  B1929+10 is
separated from its calibrator by $2^{\circ}20^{\prime}$. 

\subsection{Position determination}

Final pulsar images were made once the ionosphere calibration was
completed.  A separate image was made for each 8~MHz band.  The
position of the pulsar was measured by fitting a Gaussian ellipse to
the pulsar image.  Data from a single 8~MHz spectral window produced
images of the pulsar with signal-to-noise of about 40.  The dominant
source of the image noise was phase scatter due to the 
troposphere, making the final signal-to-noise ratio almost independent of
the pulsar's strength.  In most cases the position agreement between
images made at two different frequencies was better than 0.1~mas.  A
misalignment of more than this amount prompted a detailed inspection of the
calibration process and usually resulted in stricter elevation limits
or the removal of a frequency band with severe radio frequency
interference.  Once image alignment was achieved, a final image was
made incorporating the data from all useful frequency bands.  The
final position measurement was made from this image.  The synthesized
beam sizes for the final images were about 10~mas north--south and
4~mas east--west but varied slightly depending on source declination.

\subsection{Calculation of Parallax and Proper Motion}

We model the path of the pulsar in terms of its position at epoch
2000.0 ($\alpha_{2000}, \delta_{2000}$), proper motion ($\mu_{\alpha},
\mu_{\delta}$), and parallax ($\pi$) by
\begin{equation}
\left\{ \begin{array}{rcl}
\alpha(t) & = & \alpha_{2000} + \mu_{\alpha} \, t 
        + \pi \, f_{\alpha}(\alpha_{2000},\delta_{2000},t) \\
\delta(t) & = & \delta_{2000} + \mu_{\delta} \, t
        + \pi \, f_{\delta}(\alpha_{2000},\delta_{2000},t)
\end{array} \right.
\end{equation}
where $t$ is time since year 2000.0.  The functions $f_{\alpha}$ and
$f_{\delta}$ are the parallax displacements in right ascension 
and declination, respectively, for
an object at a distance of 1~pc as a function of the time of year.
The five parameters
are determined by minimizing the $\chi^2$ function associated with the 
above model using an iterative linear inversion.  Uncertainties in the 
fit parameters were based on the single-epoch position uncertainties.

\section{Results}

The proper motion and parallax results for the nine pulsars with new
parallaxes are shown in Table~\ref{tab:results}.  The distances to the
pulsars derived from their parallax measurement are shown in
comparison with their dispersion measure (DM) derived distances in
Table~\ref{tab:derived}.  The uncertainties in DM distance are
obtained by assigning the DM a Gaussian probability distribution
function (PDF) centered on the measured
DM with a 40\% variance.  The DM PDF provides a way to account for the
uncertainty in the dispersion model; it is not meant to reflect
uncertainty in the the dispersion measures as the DMs of the pulsars
studied in this sample are measured to better than 1\%.  Since the 
dispersion model is not smooth, a non-Gaussian distance PDF
is obtained when the dispersion
model is applied to the Gaussian DM PDF.  
The most probable distance is quoted along with errors
bracketing the most compact 68\% likelihood region.  B1237+25 has a
very high $d_{\mathrm{DM}}$ upper limit because of is to its position
high above the galactic plane ($z \sim 850$ pc).  The dispersion model
provides poor constraints to its distance since very little ionized
material is thought to be at or above this pulsar's height.  Also
listed in Table \ref{tab:derived} are the pulsars' transverse
velocities ($v_{\perp}$) and the average electron densities
($n_{\mathrm{e}}$) along lines of sight.  

The derived parameters of six other pulsars with accurate parallax
distances are shown in Table~\ref{tab:derived2}.  
Chatterjee et al. (2001) measured the parallax of B0919+06 using the
VLBA.  The proximity of an in-beam calibrator to the pulsar
($\theta_{\mathrm{sep}} < 20^{\prime}$), and their simultaneous
observation, reduce the amount of uncalibrated ionosphere and
troposphere.  Further, GPS data was used to model and remove most of
the remaining differential ionosphere.  Single epoch uncertainties as
low as 0.2~mas were attained in this manner.  The other five pulsars
are millisecond pulsars with distances determined through pulse
arrival timing.  B1534+12 has a timing parallax upper limit of $\pi <
1.8$~mas.  A more accurate distance of $d = 1.08\pm0.15$ was inferred
for this pulsar by monitoring the decrease in its binary period,
$\dot{P}_b$.  By assuming that General Relativity accurately predicts
the orbital decay of this close neutron star binary via emission of
gravitational radiation, the other contributions to $\dot{P}_b$ from
the motion of the system restricts its possible distance.  (Stairs et
al.\ 1998 \& 1999).

\begin{figure}
\plotone{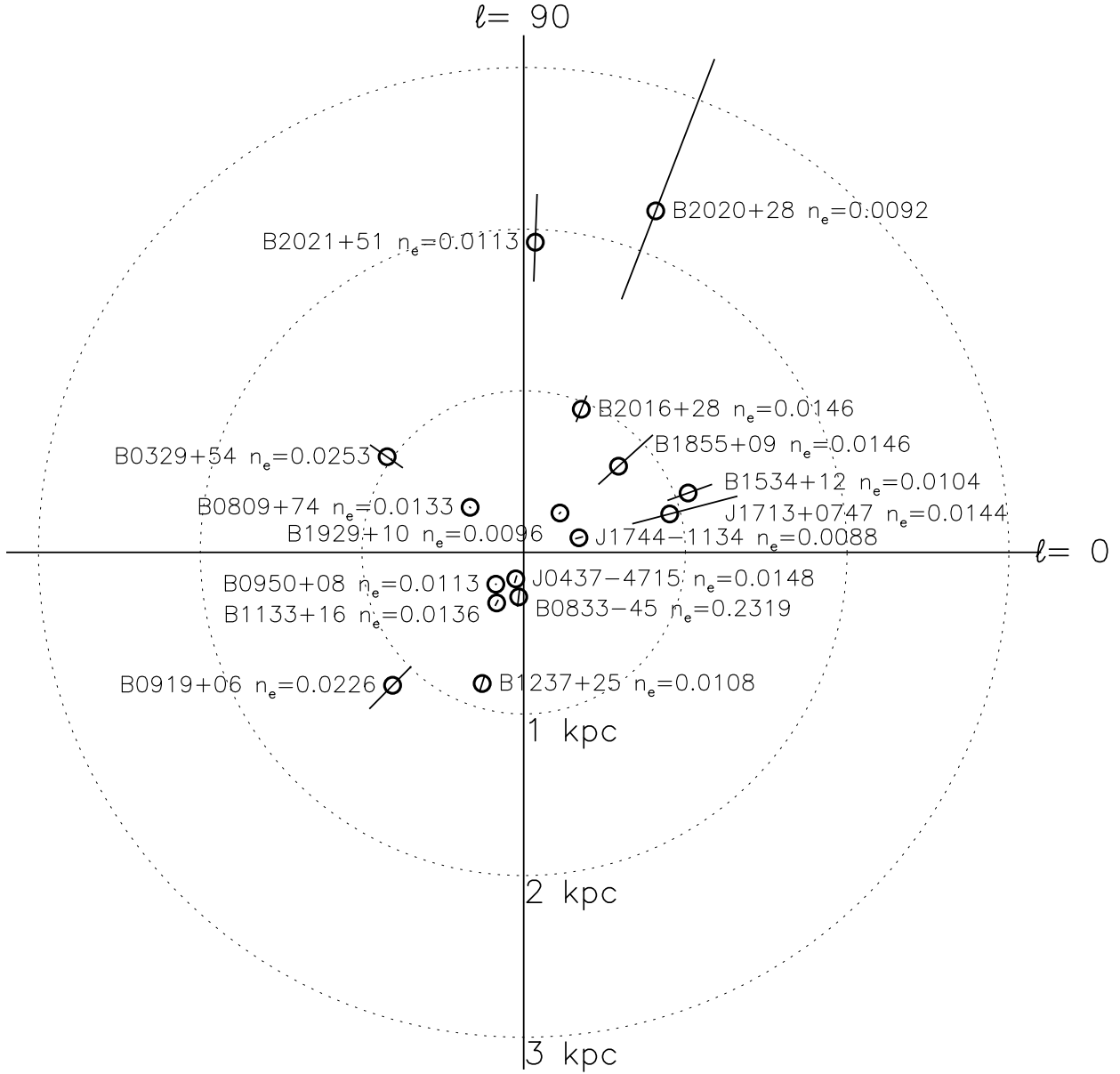}
\caption[neplot.eps]{
\label{fig:neplot}
A projection of pulsars with accurate distance measurements onto the
galactic plane.  Each pulsar is labeled with its name and estimated
line of sight electron density, $n_{\mathrm{e}}$.  
The radial line through each pulsar is its distance error bar, representing its
most compact 68\% confidence interval.
}
\end{figure}

\begin{figure}
\plotone{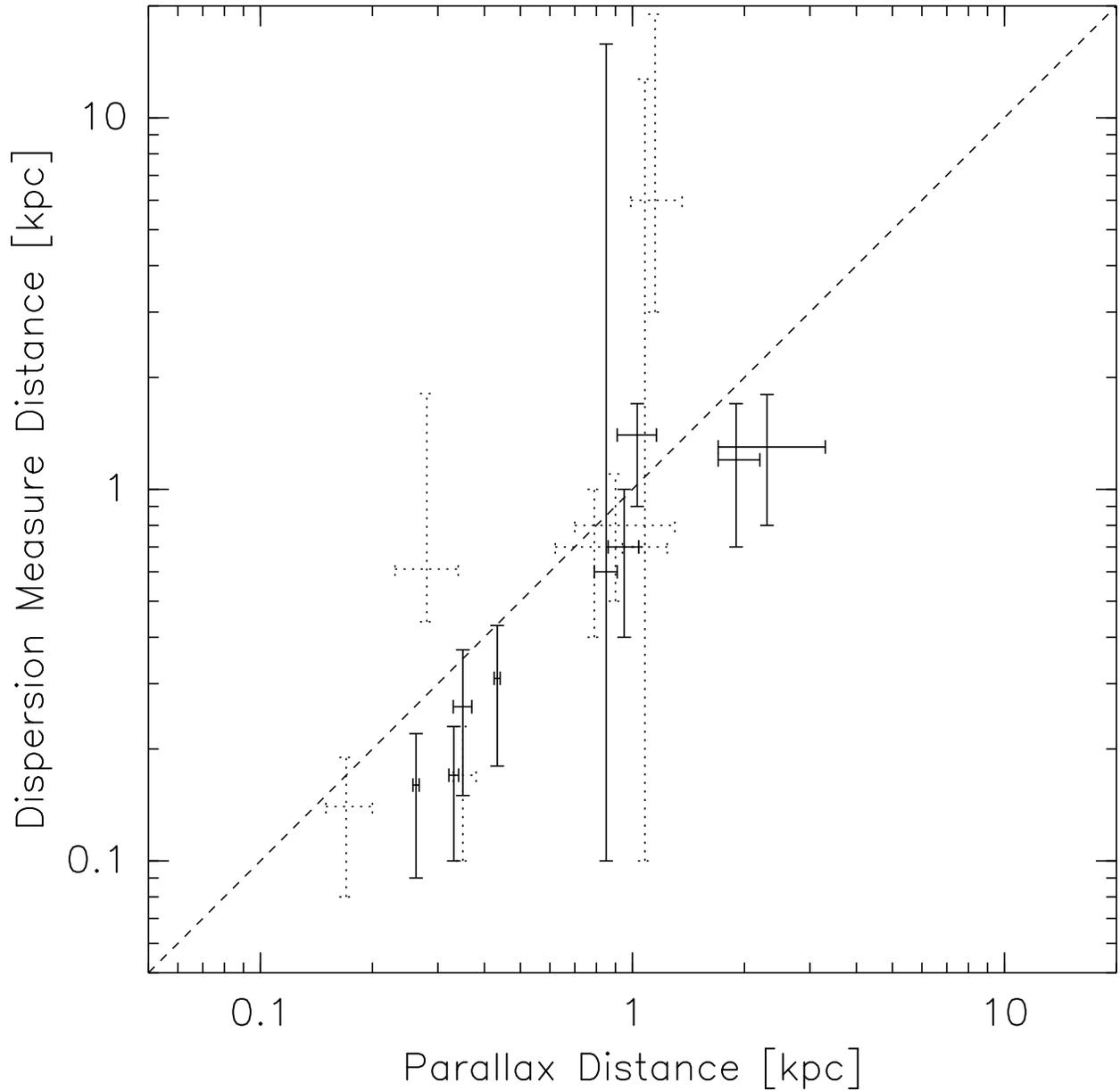}
\caption[distdist.eps]{
\label{fig:distdist}
Dispersion measure derived distances compared against parallax distances.  The
solid error bars are for the nine pulsars studied in this work (listed in
Table~\ref{tab:derived}); the dotted
error bars are for those pulsars in Table~\ref{tab:derived2}.  Error
bars in each dimension represent the most compact 68\% confidence interval.  
The predominance of pulsars to the lower right of the dashed diagonal line
suggest that the dispersion model systematically underestimates distances
to nearby pulsars.
}
\end{figure}

The distances to the pulsars in Tables~\ref{tab:derived} and
\ref{tab:derived2} are plotted in Figure~\ref{fig:neplot}.  The dispersion
measure distances to these pulsars are plotted against the parallax
distances in Figure~\ref{fig:distdist}.  The dispersion measure distances
tend to be smaller than the observed parallax distances.  This may be
due to the local low electron density bubble not being included in the
dispersion measure model.  For more detail on the local bubble, see 
Toscano et al. (1999).

\clearpage

\begin{deluxetable}{lrrrrr}
\tablewidth{0pt}
\tablecaption{Position, Proper Motion and Parallax Results \label{tab:results}}
\tablehead{
    \colhead{Pulsar}
  & \colhead{$\alpha_{2000}^{\dagger}$}
  & \colhead{$\delta_{2000}$}
  & \colhead{$\mu_{\alpha}$}
  & \colhead{$\mu_{\delta}$}
  & \colhead{$\pi$}
\\
  &
  &
  & \colhead{(mas/yr)}
  & \colhead{(mas/yr)}
  & \colhead{(mas)}
}
\startdata
B0329+54 & $03^h32^m59^s.3862$ & $54^{\circ}34'43''.5051$ & $17.00\pm0.27$ & $-9.48\pm0.37$ & $0.94\pm0.11$ \\
B0809+74 & $08^h14^m59^s.5412$ & $74^{\circ}29'05''.3671$ & $24.02\pm0.09$ & $-43.96\pm0.35$ & $2.31\pm0.04$ \\
B0950+08 & $09^h53^m09^s.3071$ & $07^{\circ}55'36''.1475$ & $-2.09\pm0.08$ & $29.46\pm0.07$ & $3.82\pm0.07$ \\
B1133+16 & $11^h36^m03^s.1829$ & $15^{\circ}51'09''.7257$ & $-73.95\pm0.38$ & $368.05\pm0.28$ & $2.80\pm0.16$ \\
B1237+25$^{\ddag}$ & $12^h39^m40^s.3589$ & $24^{\circ}53'50''.0193$ & $-106.82\pm0.17$ & $49.92\pm0.18$ & $1.16\pm0.08$ \\
B1929+10 & $19^h32^m13^s.9496$ & $10^{\circ}59'32''.4178$ & $94.82\pm0.26$ & $43.04\pm0.15$ & $3.02\pm0.09$ \\
B2016+28 & $20^h18^m03^s.8332$ & $28^{\circ}39'54''.1527$ & $-2.64\pm0.21$ & $-6.17\pm0.38$ & $1.03\pm0.10$ \\
B2020+28 & $20^h22^m37^s.0718$ & $28^{\circ}54'23''.0300$ & $-4.38\pm0.53$ & $-23.59\pm0.26$ & $0.37\pm0.12$ \\
B2021+51 & $20^h22^m49^s.8655$ & $51^{\circ}54'50''.3881$ & $-5.23\pm0.17$ & $11.54\pm0.28$ & $0.50\pm0.07$ \\
\enddata
\tablenotetext{\dagger}{A pulsar's epoch 2000.0 position, ($\alpha_{2000},
\delta_{2000}$) is based on the position of the reference source used (see
Table~\ref{tab:calibs}) and is
accurate to about 15 mas.}
\tablenotetext{\ddag}{B1237+25 data is based on its closest calibrator,
J1240+2405.}
\end{deluxetable}

\clearpage

\begin{deluxetable}{lrrrr}
\tablewidth{0pt}
\tablecaption{Derived parameters \label{tab:derived}}
\tablehead{
    \colhead{Pulsar}
  & \colhead{$d_{\mathrm{DM}}$}
  & \colhead{$d_{\pi}^{\dagger}$}
  & \colhead{$v_{\perp}$}
  & \colhead{$n_{\mathrm{e}}^{\ddag}$}
\\
  & \colhead{(kpc)}
  & \colhead{(kpc)}
  & \colhead{(km s$^{-1}$)}
  & \colhead{(cm$^{-3}$)}
}
\startdata
B0329+54 & $1.4^{+0.3}_{-0.5}$ & $1.03^{+0.13}_{-0.12}$ & $95^{+12}_{-11}$ & $0.0253\pm0.0030$ \\
B0809+74 & $0.31^{+0.12}_{-0.13}$ & $0.433^{+0.008}_{-0.008}$ & $102^{+2}_{-2}$ & $0.0133\pm0.0002$ \\
B0950+08 & $0.16^{+0.06}_{-0.07}$ & $0.262^{+0.005}_{-0.005}$ & $36.6^{+0.7}_{-0.7}$ & $0.0113\pm0.0002$ \\
B1133+16 & $0.26^{+0.11}_{-0.11}$ & $0.35^{+0.02}_{-0.02}$ & $631^{+38}_{-35}$ & $0.0136\pm0.0008$ \\
B1237+25 & $0.6^{+15.2}_{-0.5}$ & $0.85^{+0.06}_{-0.06}$ & $475^{+34}_{-30}$ & $0.0108\pm0.0007$ \\
B1929+10 & $0.17^{+0.06}_{-0.07}$ & $0.331^{+0.010}_{-0.010}$ & $163^{+4}_{-5}$ & $0.0096\pm0.0003$ \\
B2016+28 & $0.7^{+0.3}_{-0.3}$ & $0.95^{+0.09}_{-0.09}$ & $30^{+3}_{-4}$ & $0.0146\pm0.0013$ \\
B2020+28 & $1.3^{+0.5}_{-0.5}$ & $2.3^{+1.0}_{-0.6}$ & $256^{+114}_{-61}$ & $0.0092\pm0.0030$ \\
B2021+51 & $1.2^{+0.5}_{-0.5}$ & $1.9^{+0.3}_{-0.2}$ & $115^{+18}_{-15}$ & $0.0113\pm0.0016$ \\
\enddata
\tablenotetext{\dagger}{Gaussian parallax uncertainties imply non-Gaussian
distance uncertainties.  The most compact 68\% uncertainty interval is 
shown above.}
\tablenotetext{\ddag}{The mean electron density to the pulsar, $n_{\mathrm{e}}$,
is $\mathrm{DM}\,d^{-1}$.}
\end{deluxetable}

\clearpage

\begin{deluxetable}{lrrrrc}
\tablewidth{0pt}
\tablecaption{Derived parameters of other pulsars with accurate distances\label{tab:derived2}}
\tablehead{
    \colhead{Pulsar}
  & \colhead{$d_{\mathrm{DM}}$}
  & \colhead{$d_{\pi}$}
  & \colhead{$v_{\perp}$}
  & \colhead{$n_{\mathrm{e}}$}
  & \colhead{Reference}
\\
  & \colhead{(kpc)}
  & \colhead{(kpc)}
  & \colhead{(km s$^{-1}$)}
  & \colhead{(cm$^{-3}$)}
  &
}
\startdata
J0437-4715 & $0.14^{+0.05}_{-0.06}$ & $0.17^{+0.03}_{-0.02}$ & $109^{+17}_{-14}$ & $0.0148\pm0.0021$  & 1 \\
B0833-45 & $0.61^{+1.20}_{-0.17}$ & $0.28^{+0.06}_{-0.05}$ & $67^{+16}_{-12}$ & $0.2319\pm0.0477$  & 2 \\
B0919+06 & $6^{+13}_{-3}$ & $1.15^{+0.21}_{-0.16}$ & $484^{+87}_{-66}$ & $0.0226\pm0.0036$  & 3 \\
B1534+12 & $0.7^{+12.0}_{-0.6}$ & $1.08^{+0.16}_{-0.14}$ & $131^{+20}_{-17}$ & $0.0104\pm0.0014$  & 4 \\
B1855+09 & $0.7^{+0.3}_{-0.3}$ & $0.79^{+0.29}_{-0.17}$ & $23^{+8}_{-5}$ & $0.0146\pm0.0040$  & 5 \\
J1713+0747 & $0.8^{+0.3}_{-0.3}$ & $0.9^{+0.4}_{-0.2}$ & $28^{+13}_{-8}$ & $0.0144\pm0.0048$  & 6 \\
J1744-1134 & $0.17^{+0.06}_{-0.07}$ & $0.35^{+0.03}_{-0.02}$ & $35^{+2}_{-2}$ & $0.0088\pm0.0006$  & 7 \\
\enddata
\tablerefs{
(1) Sandhu et al., 1997;
(2) Caraveo et al., 2001;
(3) Chatterjee et al., 2001;
(4) Stairs et al., 1999;
(5) Kaspi et al., 1994;
(6) Camilo et al., 1994;
(7) Toscano et al., 1999
}
\end{deluxetable}

\begin{figure}
\plotone{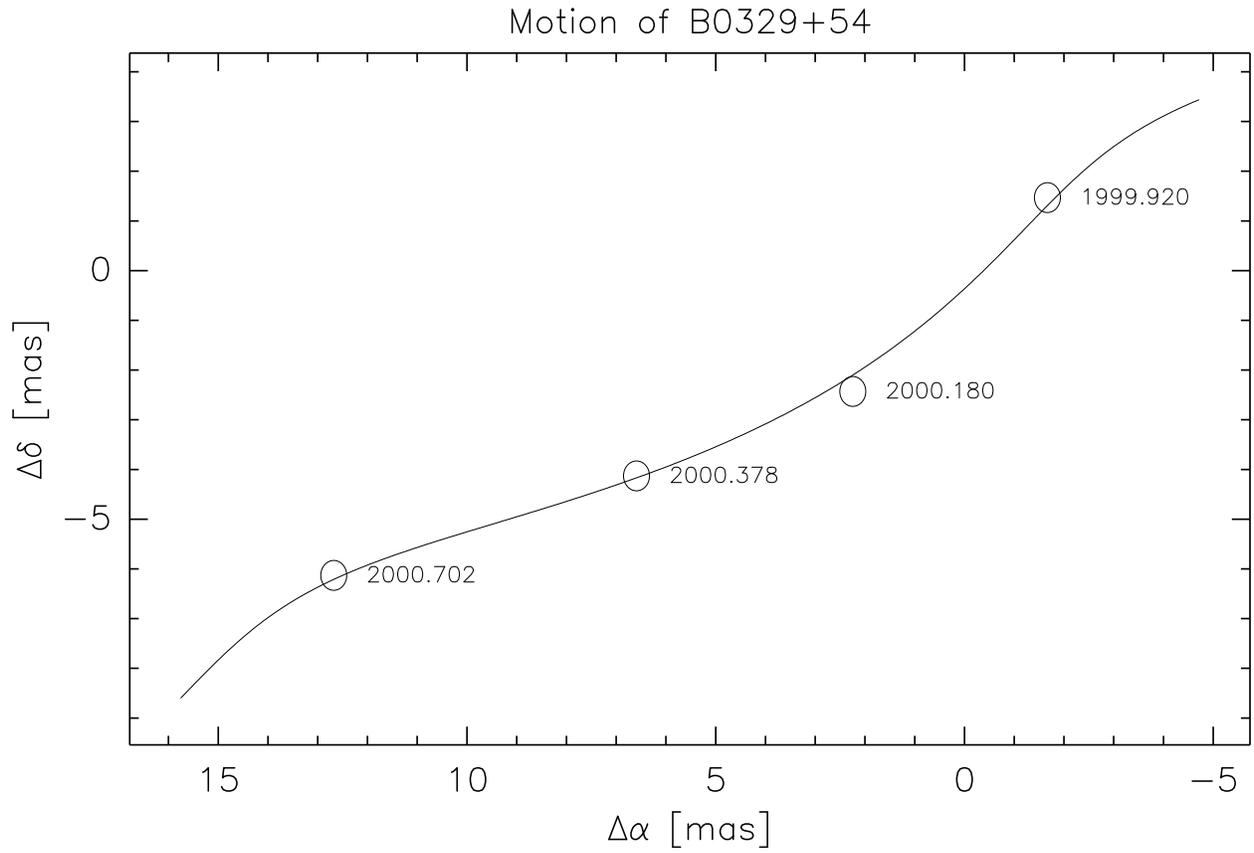}
\caption[B0329+54.eps]{
\label{fig:b0329}
The modeled path of B0329+54 based on parameter estimates (see text) 
and its four measured positions.  The
ellipses represent 1$\sigma$ position uncertainties.
}
\end{figure}

\clearpage

\clearpage

\begin{figure}
\plotone{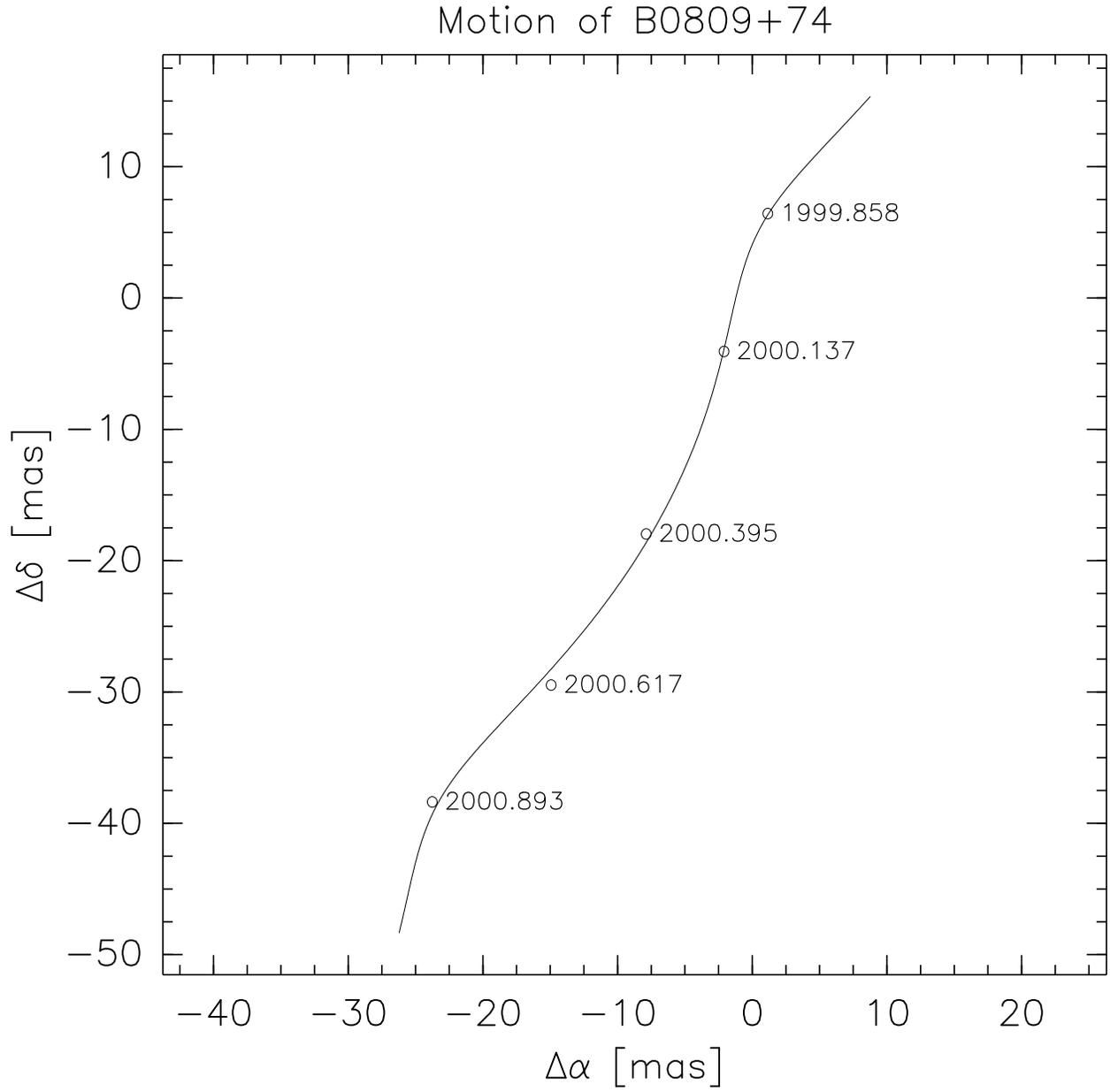}
\caption[B0809+74.eps]{
\label{fig:b0809}
The modeled path of B0809+74 and its five measured positions.
}
\end{figure}

\clearpage

\begin{figure}
\resizebox{!}{7in}{\plotone{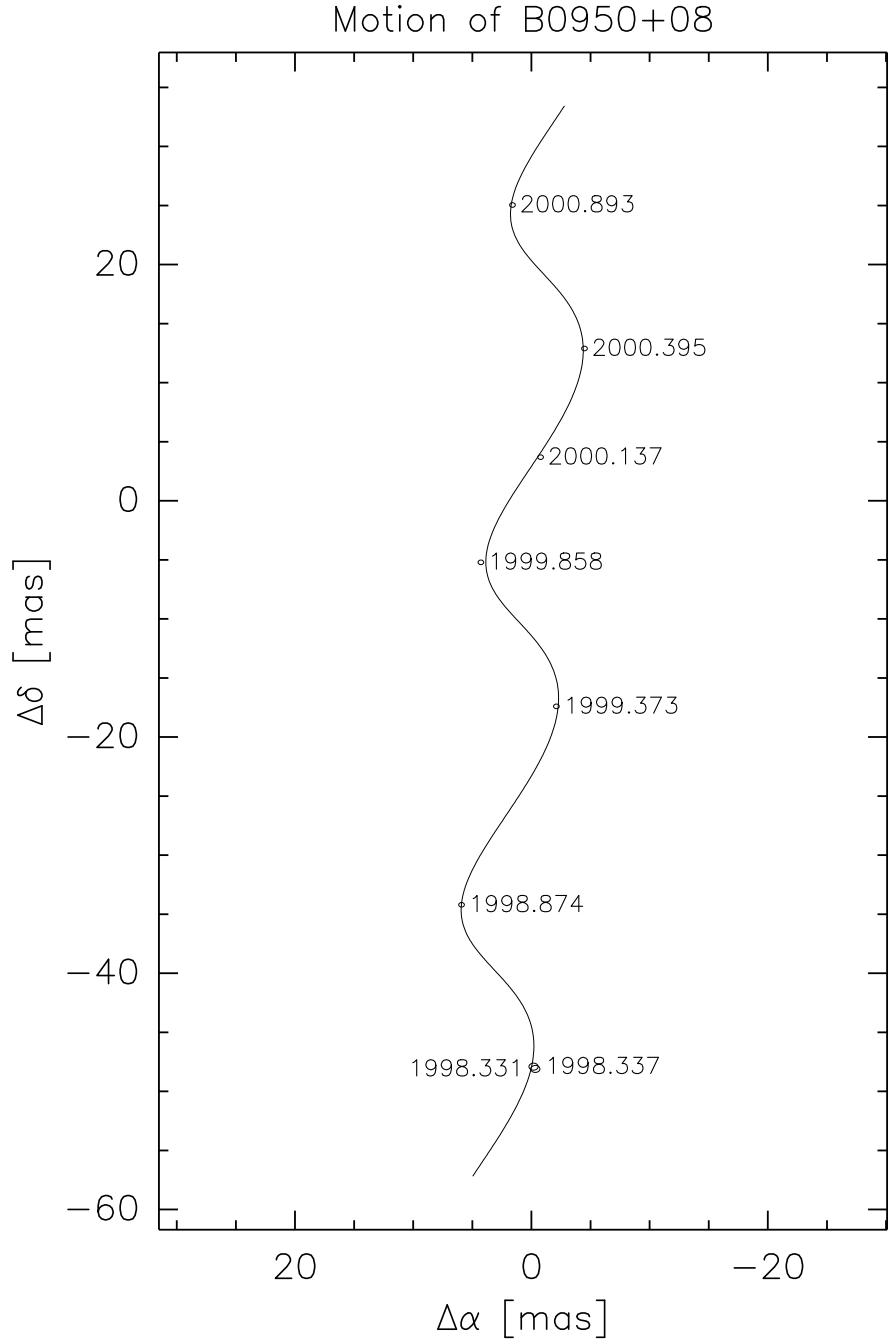}}
\caption[B0950+08.eps]{
\label{fig:b0950}
Eight observations of B0950+08 made over seven epochs and its modeled path.  
}
\end{figure}

\clearpage

\begin{figure}
\plotone{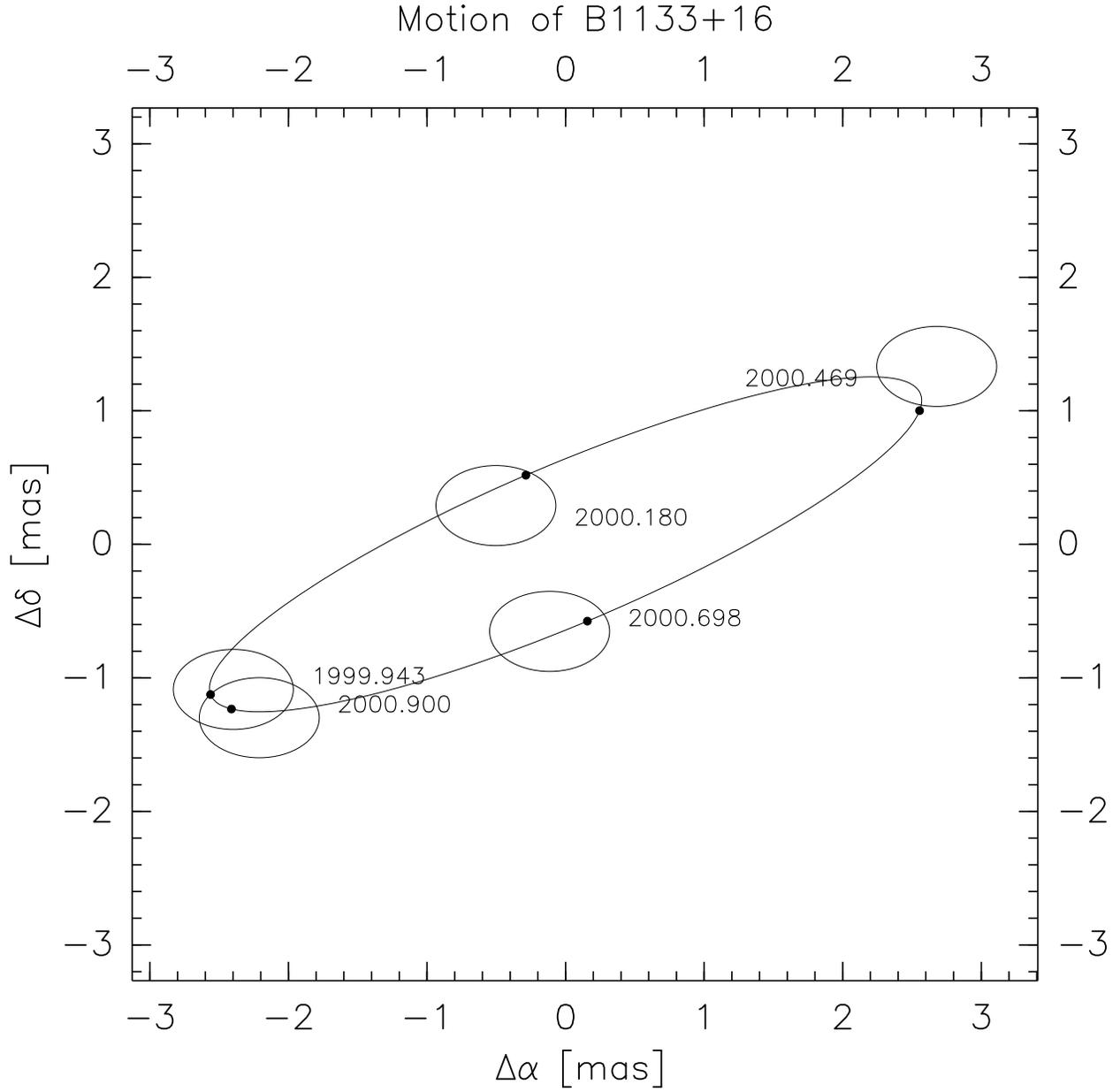}
\caption[B1133+16.eps]{
\label{fig:b1133}
A plot showing the measured positions of B1133+16 at 5 epochs and its
modeled path, both with the estimated proper motion removed.  This pulsar's
proper motion of 375~mas~yr$^{-1}$ ranks as one of the highest.  The proper
motion was subtracted to allow the significance of the parallax to be visible.
}
\end{figure}

\clearpage

\begin{figure}
\plotone{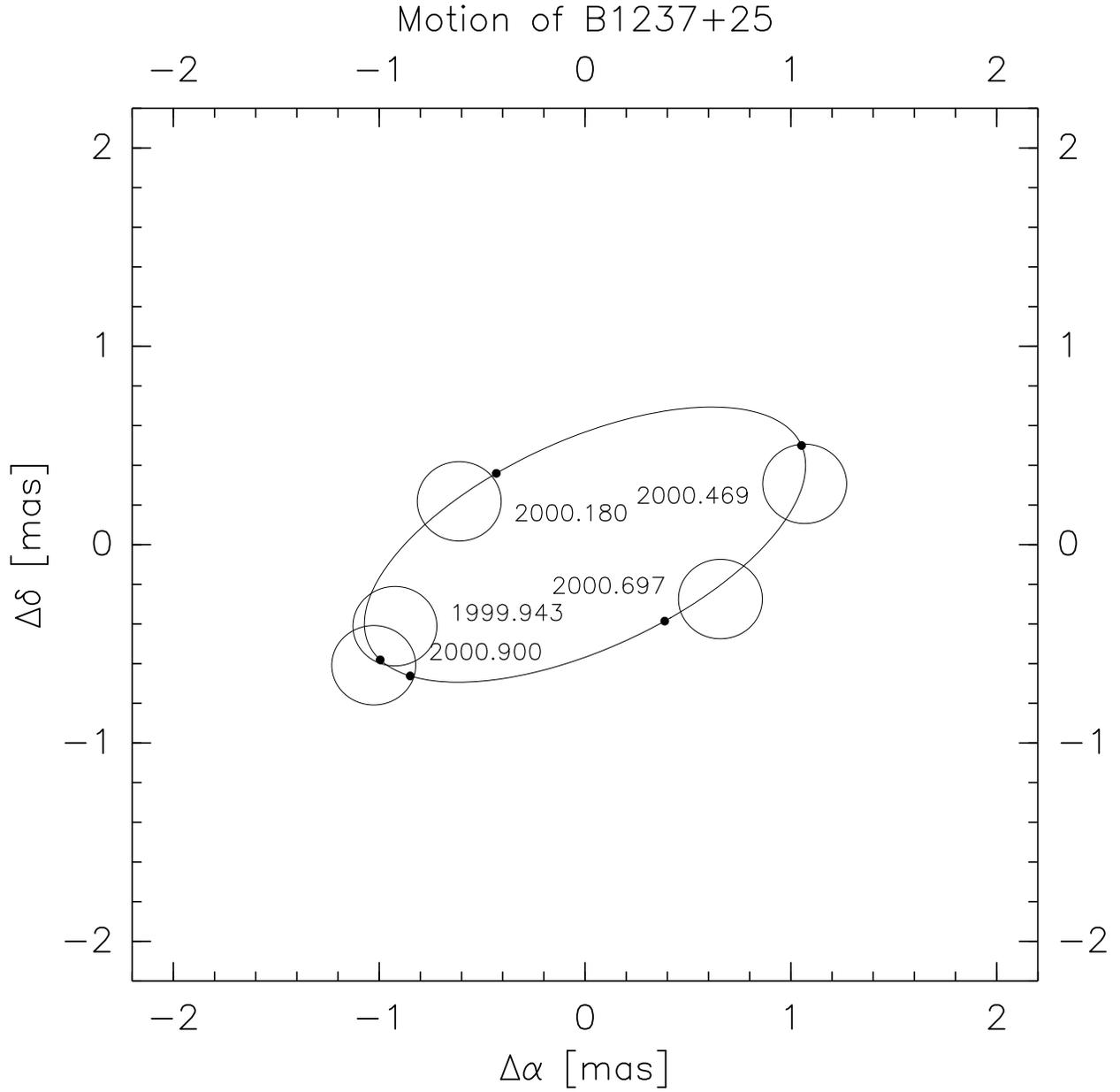}
\caption[B1237+25.eps]{
\label{fig:b1237}
A plot showing the measured positions of pulsar B1237+25 at 5 epochs
and its modeled path, both with the estimated proper motion removed.  
The linear motion of this pulsar moves it 118~mas each year.
}
\end{figure}

\clearpage

\begin{figure}
\plotone{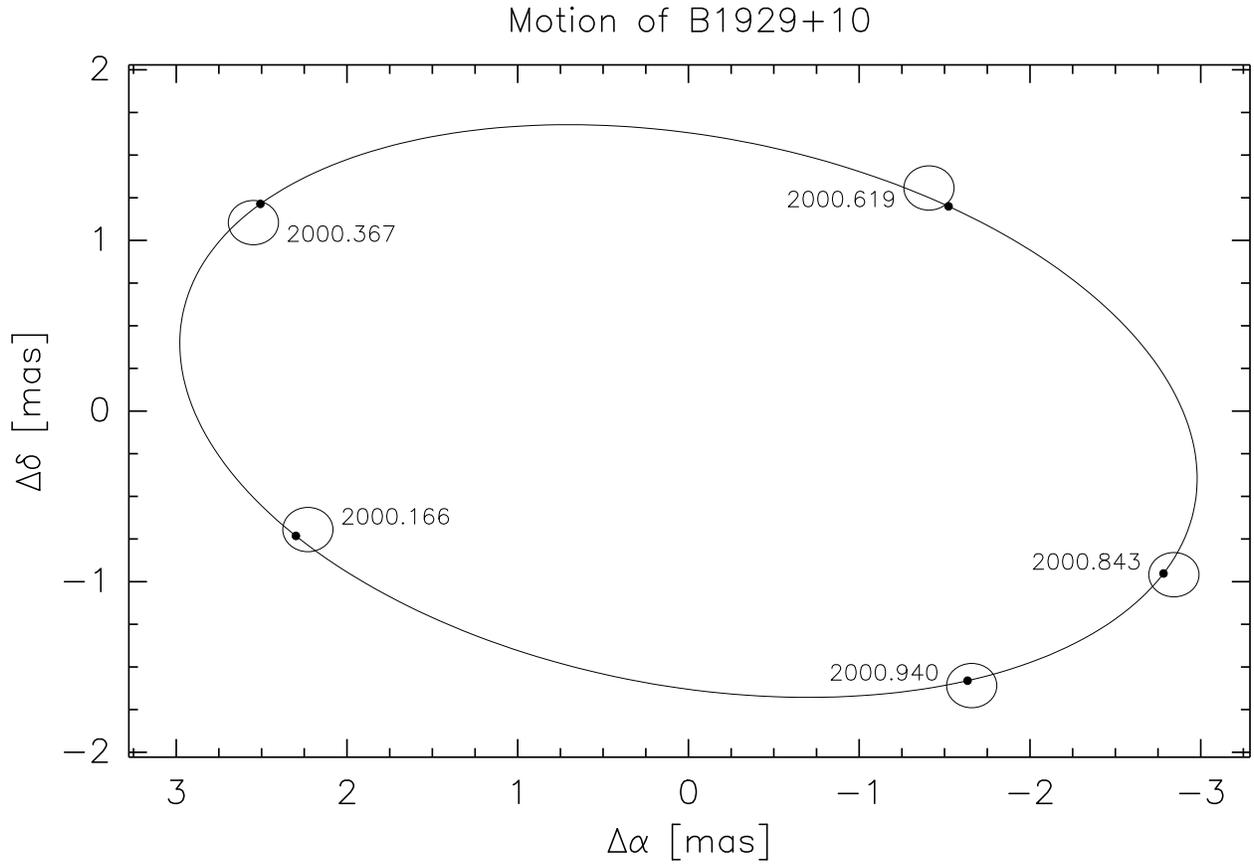}
\caption[B1929+10.eps]{
\label{fig:b1929}
A plot showing the measured positions of B1929+10 at 5 epochs and its 
modeled path, both with the estimated proper motion removed.  The proper
motion was subtracted before plotting since its linear motion over one year
amounts to 104~mas.
}
\end{figure}

\clearpage

\begin{figure}
\plotone{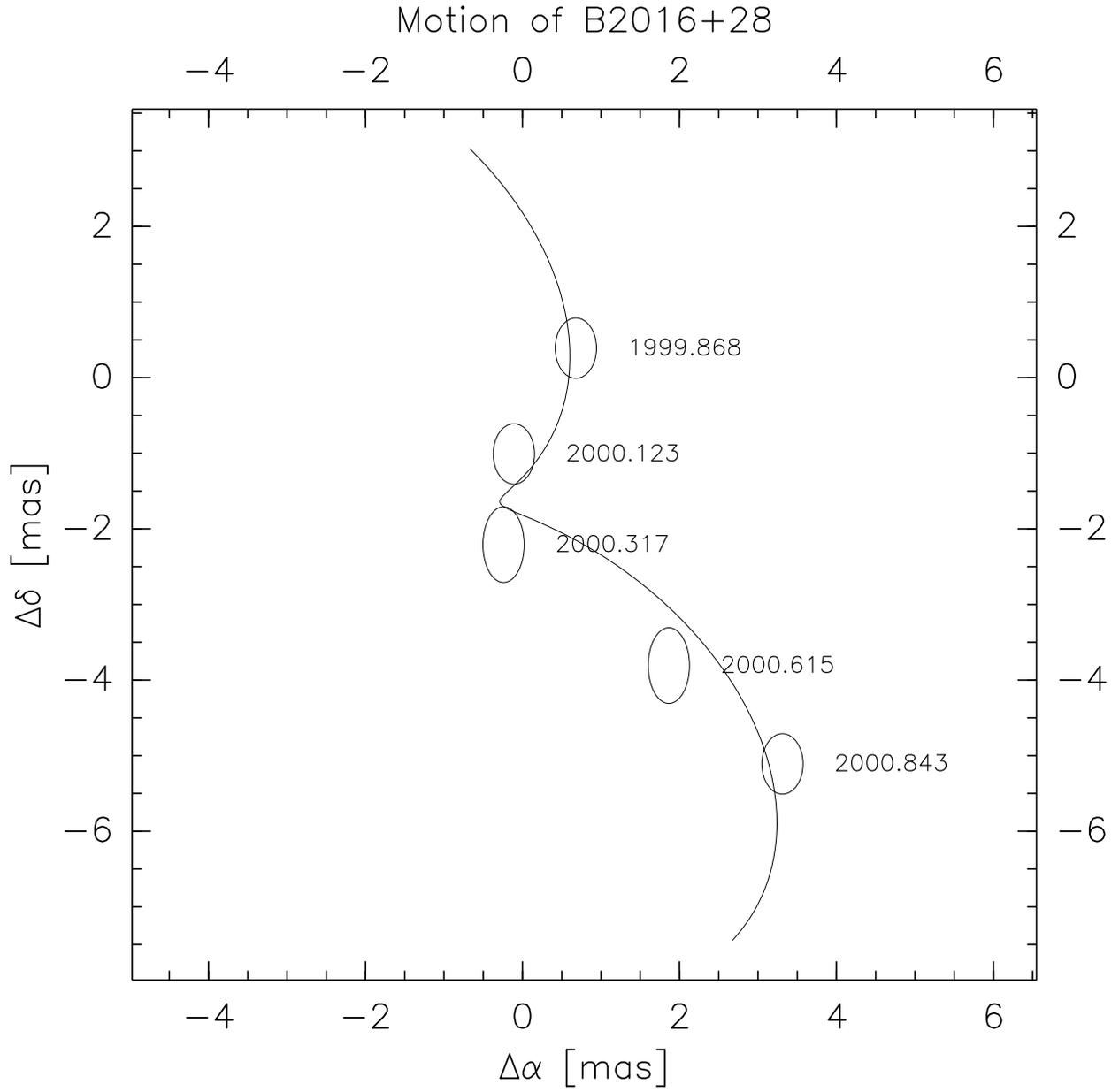}
\caption[B2016+28.eps]{
\label{fig:b2016}
The modeled path of B2016+28 and its five measured positions.
}
\end{figure}

\clearpage

\begin{figure}
\plotone{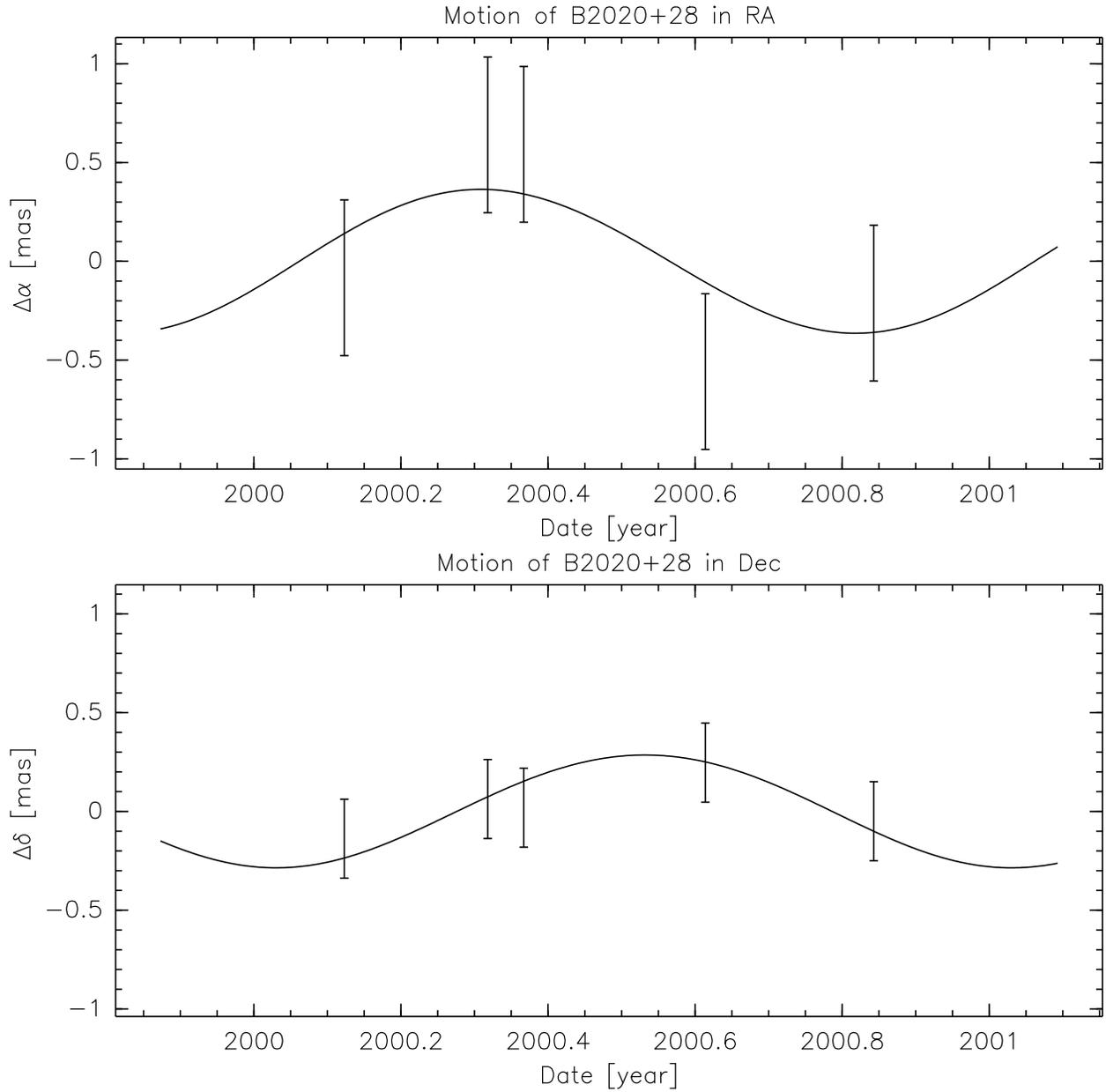}
\caption[B2020+28.eps]{
\label{fig:b2028}
The large distance to B2020+28 makes its parallax measurement the smallest of
those observed.  Its parallax fit is shown above with its proper motion 
subtracted.  Because this pulsar's single-epoch position measurements are
comparable in magnitude to the estimated parallax, the parallax fit is better 
visualized one dimension at a time.
}
\end{figure}

\clearpage

\begin{figure}
\plotone{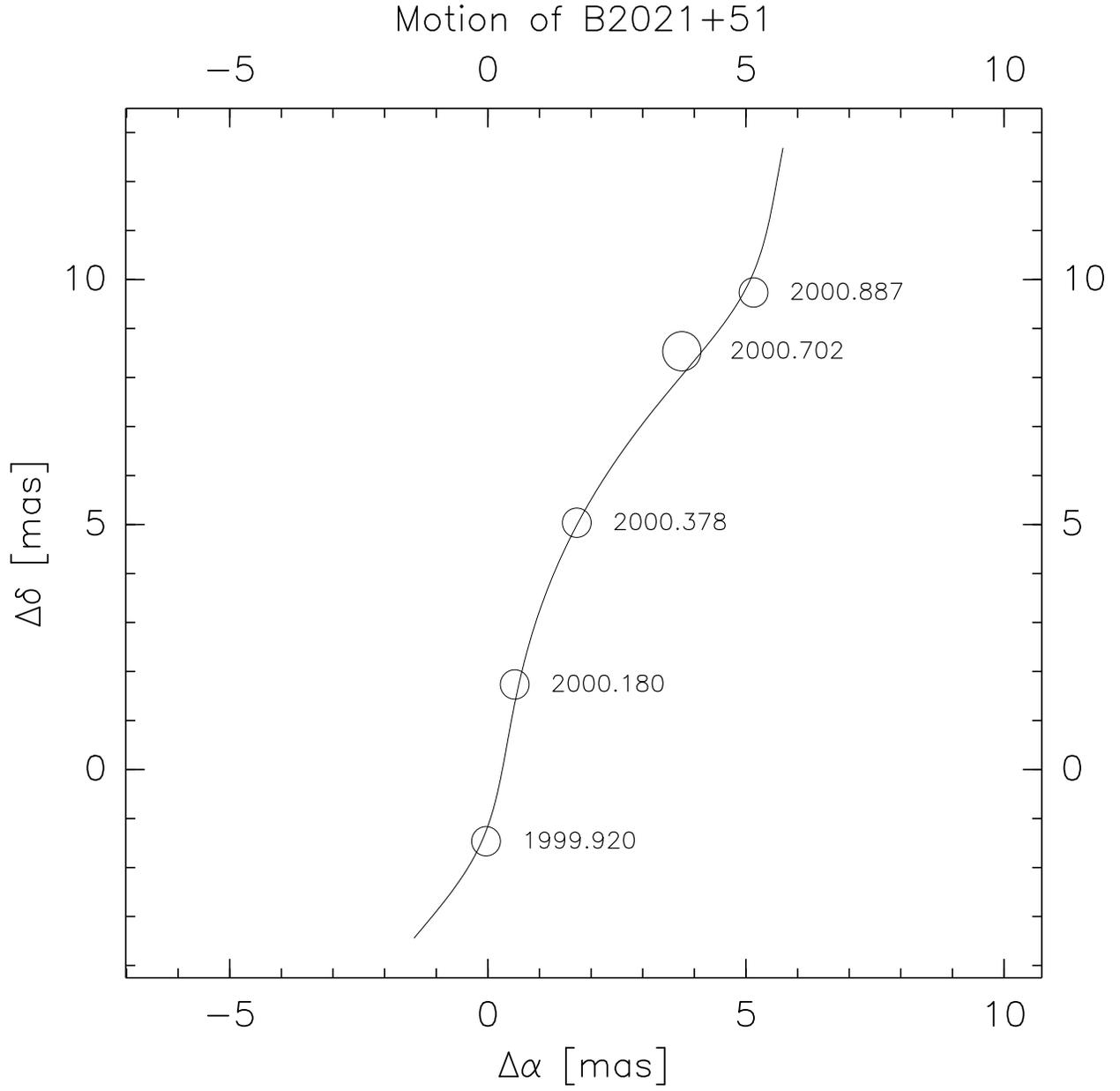}
\caption[B2021+51.eps]{
\label{fig:b2021}
The modeled path of B2021+51 and its five position measurements.
}
\end{figure}

\clearpage

\clearpage

\clearpage

\section{Uncertainties and Limitations}

Several error estimators were used to set the final parallax errors.
Pulsar images made at each of the 8 MHz bands were fit
for position.  In almost all cases scatter in position was about 0.1
mas or less.  For the brightest pulsars this scatter in position
dominated the final uncertainty.  The other single-epoch consistency
check was making images at several different short time ranges ($\sim
30$ minutes) and observing the apparent motion of the pulsar over the
$\sim 2.5$ hours of its observation.  The dominant motion was always
in right ascension, never amounting to more than 0.3~mas.  Excising
low elevation data ($< 20^{\circ}$) usually reduced this disagreement
to less than 0.1~mas.
The source of this wander is almost certainly differential
refraction in the troposphere and its effect scales with
calibrator-target separation at approximately 0.05 mas per degree of
separation, although it is very likely that the scaling is non-linear
for separations larger than those probed with these observations.
Based on the tropospheric limitations, calibrator separations greater
than the 5$^{\circ}$ limit that was initially set would probably not
generate single-epoch position measurements with accuracies better
than about 0.3 mas.  While station weather data was used to estimate 
tropospheric refraction, more sophisticated tropospheric modeling
techniques exist and should be employed in future observations.
Correlator model deficiencies, such as
incorrect Earth orientation or station positions or calibrator
positions, can also cause relative motion between the target and
calibrator.   
The success of geodesy esperiments at the 10~$\mu$as level 
using the same correlator model as was used for this work suggest that
these effects are not significant for these
observations.  The post-fit residuals provide the final check on the
uncertainties.  The least squares fit quality parameter, $\chi^2/(2
N_{\mathrm{epochs}}-5)$, was less than 1.5 for all nine pulsars,
indicating that the single epoch error estimates are reasonable.

Pulsar B1237+25 offered a unique test of the final uncertainties.  The
pulsar and two calibrators were observed together in a slightly longer
three-pointing nodding cycle.  This allowed the relative proper
motions and parallaxes of each pair to be measured.  Table
\ref{tab:1237cal} shows the proper motions and parallaxes measured for
each pair, which are in excellent agreement.

The small calibrator-to-pulsar separation is not the only limitation to the
ionosphere calibration.  The fitting of the phases imposes sensitivity
requirements.
Phase solutions accurate to $\sim 10^{\circ}$ are required on
timescales of 30~seconds or less for each frequency band.

B0950+08 now has seven epochs of VLBA astrometric data spanning just over 
2.5~years.  Figure \ref{fig:b0950} shows these measurements and its 
estimated proper motion and parallax.  
The RMS fit residual in right ascension was 0.22~mas, 
and that in declination was 0.15~mas, similar to the uncertainty estimated 
by its single-epoch position uncertainties, typically
0.2~mas.  Its distance, $262\pm5$ parsecs, is
the most accurate pulsar distance measurement made to date.  

\clearpage
\begin{deluxetable}{lrrrr}
\tablewidth{0pt}
\tablecaption{Calibration Consistency Test\label{tab:1237cal}}
\tablehead{ 
    \colhead{Calibrator}
  & \colhead{Target}
  & \colhead{$\mu_{\alpha}$}
  & \colhead{$\mu_{\delta}$}
  & \colhead{$\pi$}
\\
  & 
  & \colhead{(mas yr$^{-1}$)}
  & \colhead{(mas yr$^{-1}$)}
  & \colhead{(mas)}
}
\startdata
J1240+2405 & B1237+25   & $-106.82\pm0.17$ & $49.92\pm0.18$ & $1.16\pm0.08$ \\
J1230+2518 & B1237+25   & $-106.53\pm0.24$ & $50.21\pm0.23$ & $1.08\pm0.11$ \\
J1240+2405 & J1230+2518 & $0.10\pm0.23$ & $0.09\pm0.19$ & $0.13\pm0.11$ \\
\enddata
\end{deluxetable}
\clearpage

\section{Conclusions}

The ionosphere calibration method used in this paper successfully
removed the effect of the ionosphere from the VLBA data for all 
nine detected pulsars, allowing sub-milliarcsecond parallax 
measurements to be made.  The requirements for this technique to work
can be easily summarized -- station-based phases must be measurable
to $\sim 10^{\circ}$ at each frequency band with 30~second or shorter
solution intervals.  For our observations, the limiting flux density
at 1400~MHz is about 8~mJy.  Observing with a larger bandwidth, dividing the
bandwidth into fewer spectral windows, or adding larger VLBI stations
to the array would allow this technique to operate on weaker pulsars.
Up to 100 more pulsar parallaxes are within reach of the VLBA.

\acknowledgements

The pulsar timing data used for gating the correlator was very kindly
provided by Andrew Lyne of Jodrell Bank Observatory.  The National
Radio Astronomy Observatory is a facility of the National Science
Foundation operated under cooperative agreement by Associated
Universities, Inc.  A National Science Foundation graduate student
fellowship funded Walter Brisken.  Additional funding was provided by
Princeton University and by the NSF (AST-0098343).


\clearpage

\begin{thebibliography}{}

\bibitem[Backer \& Sramek (1982)]{bac82} Backer, D. C. \& Sramek, R. A.
        1982, \apj, 260, 512.

\bibitem[Bailes et al.(1990)]{bai90} 
Bailes, M., Manchester, R. N., Kesteven, M. J., Norris, R. P. \& Reynolds, J. E.
1990, Nature, 343, 240.

\bibitem[Beasley et al.(2002)]{bea02} Beasley, A. J., Gordon, D., Peck, A. B., 
	Petrov, L., MacMillan, D. S., Fomalont, E. B. \& Ma, C. 2002, \apjs \ 
	(accepted).

\bibitem[Brisken et al.(2000)]{bri00} Brisken, W. F., Benson, J. M.,
        Beasley, A. J., Fomalont, E. B., Goss, W. M. \& Thorsett, S. E.
        2000, \apj, 541, 959.

\bibitem[Brisken et al. (in prep)]{bri01} Brisken, W. F., Fruchter, A. S.,
Goldberg, E. E., Goss, W. M., McGary, R. S., \& Thorsett, S. E. (in prep).

\bibitem[Camilo et al.(1994)]{cam94} Camilo, F., Foster, R. S., \& Wolszczan, A.
        1994, \apj, 437L, 39.

\bibitem[Campbell et al.(1996)]{cbs+96} Campbell, R. M., Bartel, N.,
        Shapiro, I. I., Ratner, M. I., Cappallo, R. J., Whitney, A. R. \&
        Putnam, N. 1996, \apj, 461, 95L.

\bibitem[Caraveo et al. (1996)]{car96} Caraveo, P. A., Bignami, 
G. F., Mignani, R., \& Taff, L. G. 1996, \apj, 461, 91L.

\bibitem[Caraveo et al. (2001)]{cdm01} Caraveo, P. A., De Luca, A., Mignani, R. P., \& Bignami, G. F. 2001, \apj, 561, 930


\bibitem[Chatterjee et al.(2001)]{cha01} Chatterjee, S.,
        Cordes, J. M., Goss, W. M., Fomalont, E. B., Beasley, A. J.,
        Lazio, T. J. W., \& Arzoumanian, Z. 2001, \apj, 550, 287.

\bibitem[Gwinn (1986)]{gwinn} Gwinn, C. R., Taylor, J. H., Weisberg, J. M.,
        \& Rawley, L.A. 1986, AJ, 91, 338.

\bibitem[Kaspi et al.(1994)]{kas94} Kaspi, V. M., Taylor, J. H. \& Ryba, M. F.
        1994, \apj, 428, 713.




\bibitem[Salter et al.(1979)]{sal79} Salter, M. J., Lyne, A. G., \& Anderson, B.
1979, Nature, 280, 477.

\bibitem[Sandhu el al.(1997)]{san97} Sandhu, J. S., Bailes, M., 
        Manchester, R. N., Navarro, J., Kulkarni, S. R. \& Anderson, S. B. 1997,
        \apj, 478, 95L.

\bibitem[Sfeir et al.(1999)]{sfe99} Sfeir, D. M., Lallement, R., Crifo, F. \&
        Welsh, B. Y. 1999, A\&A, 346, 785.

\bibitem[Shapiro et al. (1979)]{swc+79} Shapiro, I. I., Wittels, J. J.,
Counselman, C. C., III, Robertson, D. S., Whitney, A. R., Hinteregger, H. F.,
Knight, C. A., Clark, T. A., Hutton, L. K. \& Niell, A. E. 1979, \aj, 84, 1459.

\bibitem[Stairs et al.(1998)]{sac98} Stairs, I. H., Arzoumanian, Z., 
Camilo, F., Lyne, A. G., Nice, D. J., Taylor, J. H., Thorsett, S. E., \&
Wolszczan, A. 1998, \apj, 505, 352.

\bibitem[Staris et al.(1999)]{sntt99} Stairs, I. H., Nice, D. J., Thorset, 
        S. E. \& Taylor, J. H. 1999, Gravitational Waves and Experimental Gravity, XXXIV Rencontres de Moriond.

\bibitem[Taylor \& Cordes(1993)]{tc93} Taylor, J. H., Cordes, J. M. 1993,
        \apj, 411, 674.

\bibitem[Taylor et al. (1993)]{tml93} Taylor, J. H., Manchester, R. N., \& Lyne, A. G. 1993, 
	ApJS 88, 529.

\bibitem[Taylor et al.(1999)]{tay99} Synthesis Imaging in Radio Astronomy II,
Taylor, G. B., Carilli, C. L., \& Perley, R. A., 1998, ASP Conference Series,
Vol. 180. 1999.

\bibitem[Toscano et al.(1999)]{tos99} Toscano, M., Britton, M. C., 
        Manchester, R. N., Bailes, M., Sandhu, S. R., Kulkarni, S. R. \&
        Anderson, S. B. 1999, ApJL,
        523, 171.

\bibitem[Walter (2001)]{wal01} Walter, F. M. 2001, \apj, 549, 433.

\bibitem[Weisberg et al.(1980)]{wei80} Weisberg, J. M., Rankin, J.,
        \& Boriakoff, V. 1980, A\&A, 88, 84.

\end{thebibliography}
\end{document}